\documentclass[manuscript]{aastex}
\usepackage{latexsym,epsfig,graphics,graphicx,fullpage,amsmath}
\def \be{\begin{equation}}
\def \ee{\end{equation}}
\def \bea{\begin{eqnarray}}
\def \eea{\end{eqnarray}}

\begin{document}

\title{Theoretical Estimates of 2-point Shear Correlation Functions Using Tangled Magnetic Field Power Spectrum}
\author{Kanhaiya L. Pandey\altaffilmark{1} AND Shiv K. Sethi\altaffilmark{1}}
\altaffiltext{1}{Raman Research Institute, Bangalore 560080, India}

\begin{abstract}
The existence of primordial magnetic fields can induce matter perturbations
with additional power at small scales 
 as compared to the usual $\Lambda$CDM model. We study its
 implication within the context of two-point shear correlation function from 
gravitational lensing.  We show that primordial 
magnetic field can leave its imprints on the shear correlation function
at angular scales $\lesssim \hbox{a few arcmin}$. The results 
are compared with  CFHTLS  data, which yields some of 
the strongest known constraints on the parameters (strength 
and spectral index)  of the 
primordial magnetic field.  We also discuss the possibility of detecting 
sub-nano Gauss fields using future missions such as SNAP. 
\end{abstract}


\section{Introduction}

In recent years, weak gravitational lensing has proved to be one of 
best probes of the matter power spectrum of the universe. In particular, 
this method  can reliably estimate the matter power spectrum at  small scales which
are not directly accessible to other methods e.g. galaxy surveys (for details
and further references see e.g. \citet{Munshi08,Hoekstra08,Refreigier03,BS01}).

Magnetic fields play an important role in the many areas of astrophysics, 
and are ubiquitously seen in 
the universe. They have been observed in the galaxies and clusters of
 galaxies with the coherence lengths up to 
$\simeq$ 10\hbox{--}100 kpc (for a review see e.g. \citet{Widrow02}). There is
 also  evidence of coherent magnetic fields up to super-cluster 
scales \citep{Kim89}. Still little  is known about the origin of 
 cosmic magnetic fields,  and their role in the evolutionary history of the universe. These
fields could have originated from  dynamo amplification of very tiny
 seed magnetic fields $\simeq 10^{-20} \, \rm G$ (e.g  \citet{Parker79,ZRS83,RSS88}).  It has been shown that dynamo mechanism can amplify 
fields to significant values in collapsing objects at high redshifts 
\citep{Ryu08,Schleicher10,Arshakian09,dSouza10,Federrath11a,Federrath11b,Schober11}.   It is also possible 
that much larger primordial magnetic field ($\simeq 10^{-9} \, \rm G$)  were generated during the 
inflationary phase \citep{TW88,Ratra92} and the large scale magnetic field
observed today are the relics of these fields.  In the latter case, of interest to us in this paper,  magnetic field starts with a large value in the intergalactic medium, while in the former case large magnetic fields are confined 
to bound objects.

While the presence of primordial magnetic fields have  the potential
 to explain the 
observed magnetic fields coherent at a range of scales in the present universe, 
such fields also leave detectable  signatures in important observables
at cosmological scales in the universe.

The impact  of large-scale primordial magnetic fields on CMBR temperature 
and polarization anisotropies has been studied in detail (e.g. \citet{Kandu98b, Kandu02,Sesh01,Mack02,Lewis04,GS05,TS06,Sethi05,SBK08,Sethi09,Sethi10,Tina05,Giovannini08,Yamazaki08,Sesh09}).  More recently, lower bounds $\simeq 10^{-15} \, \rm G$ on the strength of magnetic fields  have been
obtained based on observations of high-energy $\gamma$-ray photons (e.g. 
\citet{Dolag10,NV2010,Tavecchio10,Taylor11}).

\citet{Wasserman78} showed that primordial magnetic fields can
induce density perturbations in the post-recombination universe. Further
 work along these lines
have investigated  the impact of this effect for the formation of first structures, reionization of the universe, and the signal from redshifted HI line from the epoch of reionization \citep[e.g.][]{Kim96,GS03,Sethi05,TS06,sbk,Sethi09}. 
 The matter power spectrum induced by
primordial magnetic fields can dominate the matter power spectrum of  
the standard $\Lambda$CDM model at small scales. Weak gravitational lensing 
can directly probe this difference and therefore reveal the presence of primordial fields or put additional constraint on their strength.

In this paper we  attempt to constrain  primordial magnetic 
fields  within the framework of  the 
 two-point shear correlation function induced by gravitational lensing, 
 including the contribution of 
matter perturbations induced by 
 these  magnetic fields. We compare our results with the 
 CFHTLS Wide data  \citep{LFu08}.

Throughout the paper,  we used flat (k=0) $\Lambda$CDM universe with $\Omega_m$ = 0.24, $\Omega_b$ = 0.044, $h$ = 0.73 and 
$\sigma_8$ = 0.77.
\\

\section{Matter  Power Spectrum}

Tangled magnetic fields can be characterized by a power-law power spectrum:
$M(k) = A k^n$. In the pre-recombination era, the magnetic fields
are dissipated at scales below a scale corresponding to $k_{\rm max} \simeq 200 \times (10^{-9} \, {\rm G}/B_{\rm eff})$ \citep[e.g.][]{jedam98,kandu98a}. Here $B_{\rm eff}$ is the RMS at this cut-off
scale for a given value of the spectral index, $n$. Another possible normalization, commonly used in the
 literature, is the value of RMS at $k = 1 \, \rm Mpc^{-1}$, $B_0$. 
These two normalizations are related as: $B_{\rm eff} = B_0 k_{\rm max}^{(n+3)/2}$. It is possible to present results using either of the pair
 $\{B_{\rm eff},n\}$ or  $\{B_0,n\}$.  

Tangled Magnetic fields induce matter perturbations in the post-recombination
era which grow by gravitational collapse. The matter power spectrum of these
perturbations is given by: $P(k) \propto k^{2n+7}$, for $n < -1.5$, the range 
of spectral indices we consider here \citep{Wasserman78,Kim96,GS03}.  

The Magnetic field induced matter power spectrum  is cut-off at
the magnetic field  Jeans' wave number: $k_J \simeq 15 (10^{-9} \, \rm G/B_{\rm eff})$ \citep[e.g.][]{Kim96,Tina2011}.
The dissipation of tangled magnetic field in the post-recombination era also
results in an increase in the thermal Jeans' length \citep{Sethi05,SBK08}. For 
most of the range of magnetic field strengths considered here,
the  scale corresponding to $k_J$ generally exceed or are comparable to the 
thermal Jeans length (Figure~4 of \citet{SBK08}).  

For our computation, we need to know the  time evolution of the matter power
spectrum induced by tangled magnetic fields. It can be shown that the dominant
 growing mode in this case  has the same time dependence as the $\Lambda$CDM model (see e.g. \citet{GS03} and references therein)

\section{Weak Lensing \& Cosmic Shear}
The cosmic shear power spectrum $P_{k}(\ell)$ or the lensing convergence power spectrum, $P_\kappa$, is
the measure  of the projection of matter power spectrum  $P_\delta$ and
 is given by the following expression 
(\citet{BS01}),
\bea \label{eq:shearps}
P_{\kappa}(\ell)& = &  \frac {9}{4} {\Omega_m^2} \left(\frac {H_0}{c} \right) ^{4} \int_0^{\chi_{lim}} 
                    \frac {d\chi} {a^2 (\chi )} P_\delta \left( \frac {\ell}{f_K(\chi)}
                    ;\chi \right)   \nonumber\\[2ex]
                   && \times \left[ \int _{\chi} ^{\chi_{lim}} d \chi ' n(\chi ') \frac 
                    {f_K(\chi ' - \chi)} {f_K(\chi ')} \right] ^2
\eea
where $\chi$ is the comoving distance along the light ray and $\chi_{lim}$ is the limiting comoving 
distance of the survey; $f_{K}(\chi)$ is the comoving angular diameter distance; for spatially flat (K=0) universe $f_K (\chi)$ is 
numerically equal to the $\chi$ and the expression for $\chi$ in the flat universe is as given below, 

\be \label{eq:chi_z}
\chi(z) = \frac {c} {H_0} \int_0^z (\Omega_m (1+z)^3 + \Omega_{\Lambda})^{-1/2} dz
\ee
n(z) is the redshift distribution of the sources and $\ell$ is the modulus of a two dimensional wave vector perpendicular 
to the line of sight. $P_\delta$ is the matter power spectrum. In this 
paper, we use tangled magnetic power spectrum as $P_\delta$ to compute 
the shear power spectrum  for the magnetic cases.

The cosmological shear field induced by 
density perturbations  is a curl-free  quantity and is donated as an 
E-type field. One  can decompose the observed shear signal into E (non-rotational) and 
B (rotational) components.  Detection of non-zero B-modes indicates a
 non-gravitational 
contribution to the shear field, which might be caused by systematic contamination to the lensing signal.\footnote{The presence of primordial magnetic fields
will also generate the B-modes of the shear power spectrum. Both the vector
and tensor modes generated by magnetic fields could  sources these modes. Vector
modes are likely to play a more dominant role at angular scales of interest to
us in the paper. We hope to  explore this possibility in a future work.}

These decomposed shear correlation functions can be expressed as:
\be \label{eq:xieb}
\xi_{E,B}(\theta)=\frac {\xi_+(\theta) \pm \xi'(\theta)} {2}
\ee
where $\xi'$ is given by
\be \label{eq:xid}
\xi'(\theta) = \xi_-(\theta) + \int_\theta^\infty \frac { d \vartheta} {\vartheta} \xi_-(\vartheta) \left(
       4-12 \left(\frac {\theta} {\vartheta} \right)^2 \right)
\ee
$\xi_+$ and $\xi_-$ are  two-point shear correlation functions which are 
 related to the matter  power 
spectrum according to the following relation, 
\be \label{eq:xi_pm}
\xi_\pm(\theta) = \frac {1} {2\pi} \int_0^\infty d \ell \ \ell P_\kappa (\ell) J_{0,4} (\ell \theta)
\ee
$\theta$ is the angular separation between the galaxy pairs, and $J_{0,4}$ are Bessel functions of the first  kind.

\section{Shear power spectrum  from  tangled magnetic field power spectrum}

We use  the tangled magnetic field matter 
 power spectrum   $P_\delta$  to compute the 
 shear power spectrum $P_\kappa (\ell)$ which in turn is used 
to calculate $\xi_+$, $\xi_-$, $\xi_E$ and $\xi_B$ using Eqs~\eqref{eq:xieb}, \eqref{eq:xid} \& \eqref{eq:xi_pm}.
 We have used the same source redshift 
distribution as  in \citet{LFu08}:
\be \label{eq:n_z}
n(z) = A \frac {z^a + z^{ab}} {z^b + c} ; \ \ A = \left( \int_0^{z_{max}} \frac {z^a + z^{ab}} {z^b + c} dz \right)^{-1}
\ee
where $z_{max}$ = 6. Values of the parameters a, b, c \& A we have taken from the same paper \citet{LFu08}. Values 
of these parameters as quoted in the paper are as, a = 0.612 $\pm$ 0.043 ; b = 8.125 $\pm$ 0.871 ; c = 0.620 $\pm$ 0.065 
\& A = 1.555. To evaluate the integral \eqref{eq:shearps} we changed the variable from $\chi$ to z using \eqref{eq:chi_z}.
\bea \label{eq:shearpsz}
P_{\kappa}(\ell)& = &  \frac {9}{4} {\Omega_m^2} \left(\frac {H_0}{c} \right) ^{4} \int_0^{z_{lim}} 
                    \frac {dz} {a^2 (z)} P_\delta \left(k,z \right)   \nonumber\\[2ex]
                   && \times \left[ \int_z^{z_{lim}} d z' \ n(z') \ \frac {\chi (z' - z)} {\chi (z')} \right ] ^2
\eea
where $k = \ell / \chi (z)$. again $P_{\delta}$ (k,z) can be written as,
\be \label{eq:p_kz}
P_{\delta} (k,z) = P_{\delta} (k) \times D^2 (z)
\ee
where D(z) is growth factor, which as noted above 
is the same as for the flat $\Lambda$CDM mode and is  given by \citet{Peebles93}:
\be \label{eq:D_z}
D(z) = \frac {5\Omega_m}{2}  [\Omega_m (1+z)^3 + \Omega_\Lambda ]^{1/2} \int_z^{\infty} \frac {1+z} {(\Omega_m (1+z)^3 +
        \Omega_\Lambda)^{3/2}} \ dz
\ee
We took $z_{lim}$ = 2.5 for our calculations as in \citet{LFu08}.

For comparison, we also compute all the relevant quantities
for the  linear and non-linear  $\Lambda$CDM  models. For
 $\Lambda$CDM linear power 
spectrum we used $P(k,z) = A k T^2(k) D^2(z)$, where the  transfer 
function  $T(k)$ is given by \citet{BE84}. For nonlinear 
$\Lambda$CDM we followed prescription given in \citet{PD96}.

\begin{figure}
   \centerline{
   \includegraphics[scale=0.7,angle=0]{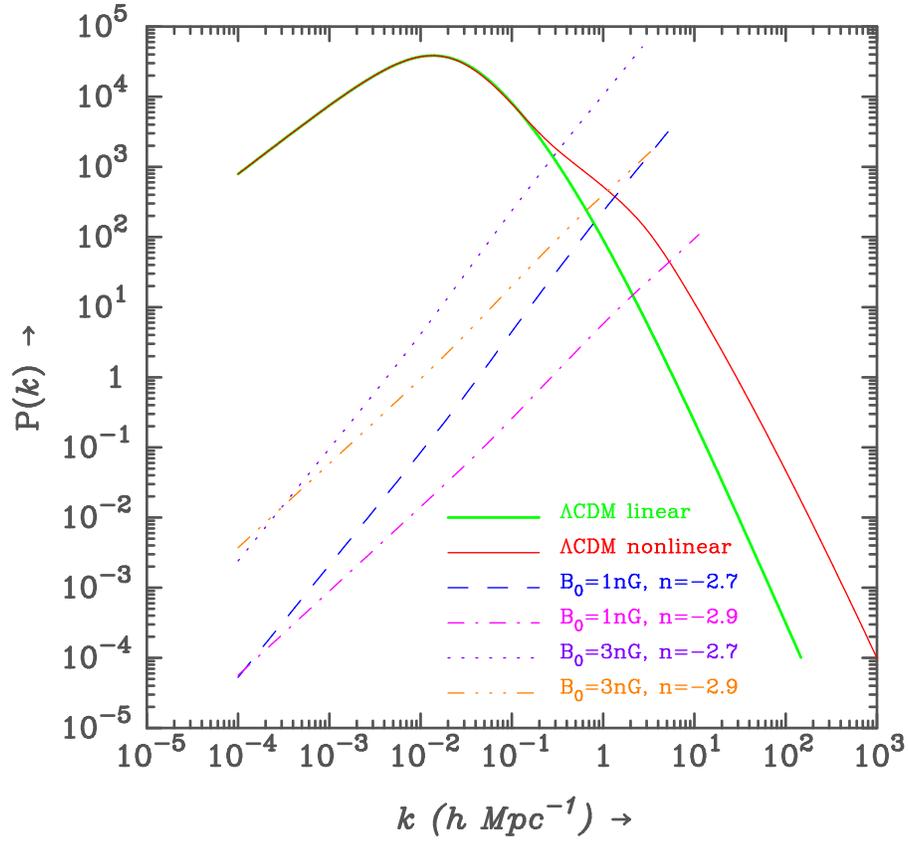}
}
   \caption[]{
The Matter power spectrum is displayed for the magnetic and non magnetic cases.
Magnetic field-induced matter power spectra are plotted for $k < k_{\rm J}$
in each case. 
}
\end{figure}
\vspace{0.1in}
\section{Results}
In Figure~1 we show the tangled  magnetic field matter
 power spectra for a range of spectral indices $n$ and magnetic 
field strengths, $B_0$ at $z = 0$. 
 The matter  power spectra
are plotted for $k < k_J$; a sharp cut-off below this scale is 
assumed for our computation. For comparison,   we have also displayed 
 the linear and non-linear $\Lambda$CDM  matter power spectra
(the non-linear power spectrum is obtained following  the method
introduced by  \citet{PD96}). The figure shows that
the magnetic field induced matter power spectra can dominate over
the $\Lambda$CDM case at small scales. Possible implications of 
this excess have already been studied for early formation of structures, 
reionization, and the HI signal from the epoch of reionization \citep{Sethi05,TS06,sbk,Sethi09,Sethi10}. Here 
we explore the observational signatures of this excess in the weak lensing 
data. 

In Figure~2 we show the shear power spectra for magnetic and non-magnetic cases. The green and  red curves present the shear power 
spectrum for $\Lambda$CDM linear and nonlinear matter power spectra, respectively. The blue curve shows the shear power spectrum for the 
tangled magnetic field power spectrum ($B_{\rm eff}$ = 3.0 nG and  $n$ = -2.9). In
 this figure  we can see the impact of additional power in the  
tangled magnetic field-induced matter 
power spectrum as an enhancement in the shear power spectrum
  on  angular scales $\simeq 1'$. 

 The peak of the matter  power spectra of 
both the $\Lambda$CDM model and the magnetic-field induced matter power 
spectra are also seen in the shear power spectra. The ratio of angular 
scales at the peak of the two cases correspond to the ratio
of these peaks of the matter power spectra:  $k_{\rm eq}/k_J$. In the $\Lambda$CDM model the power at small scales falls as $k^{-3}$, while $k_J$ imposes
a sharp cut-off in the magnetic case. In both the cases, there is power
at angular scales smaller than the peak of the matter power spectra. But 
the sharp cut-off in the matter power spectrum at $k > k_J$ 
 results in  a  steeper drop in shear power spectra as compared
to the $\Lambda$CDM case. This cut-off ensures that
the magnetic field-induced effects dominate the shear power spectrum for only
a small range of angular scales.

In Figure~3,  the two-point shear correlation functions $\xi_E$
and $\xi_B$  are shown  for magnetic and non-magnetic cases. As noted in the 
previous section, we use the parameters of the paper of  \citet{LFu08}
for all our computation, which allows us to directly compare our 
results with their data, shown in Figure~3. 

For detailed comparison with 
the data of \citet{LFu08}, we performed a  $\chi^2$ including 
the effect of both the $\Lambda$CDM (non-linear model with the best
fit parameters as obtained by \citet{LFu08}) and the magnetic field 
induced signal. We fitted the sum of these two signals ($({\xi_E})_B$ + $({\xi_E})_{\Lambda CDM}$) against the CFHTLS data to obtain limits on the 
magnetic field strength $B_0$ and the spectral index $n$. As seen in Figure~3, 
the magnetic field induced signal dominates the data for only a small range of 
angular scales below a few arc-minutes. However, this can put stringent 
constraints on the magnetic field model. Our best fit values are
$B_0 = 1.5 \, \rm  nG$ and $n= -2.96$. In Figure~4, we show the allowed contours
of these parameters for a range of  $\Delta\chi^2 = \chi^2_i - \chi^2_{min}$.
It should be noted that $B_0 =0$ is an acceptable fit to the data because 
we fix  the best fit parameters obtained by \citet{LFu08}. 
\vspace{0.1in}
\begin{figure}
   \centerline{
   \includegraphics[scale=0.7,angle=0]{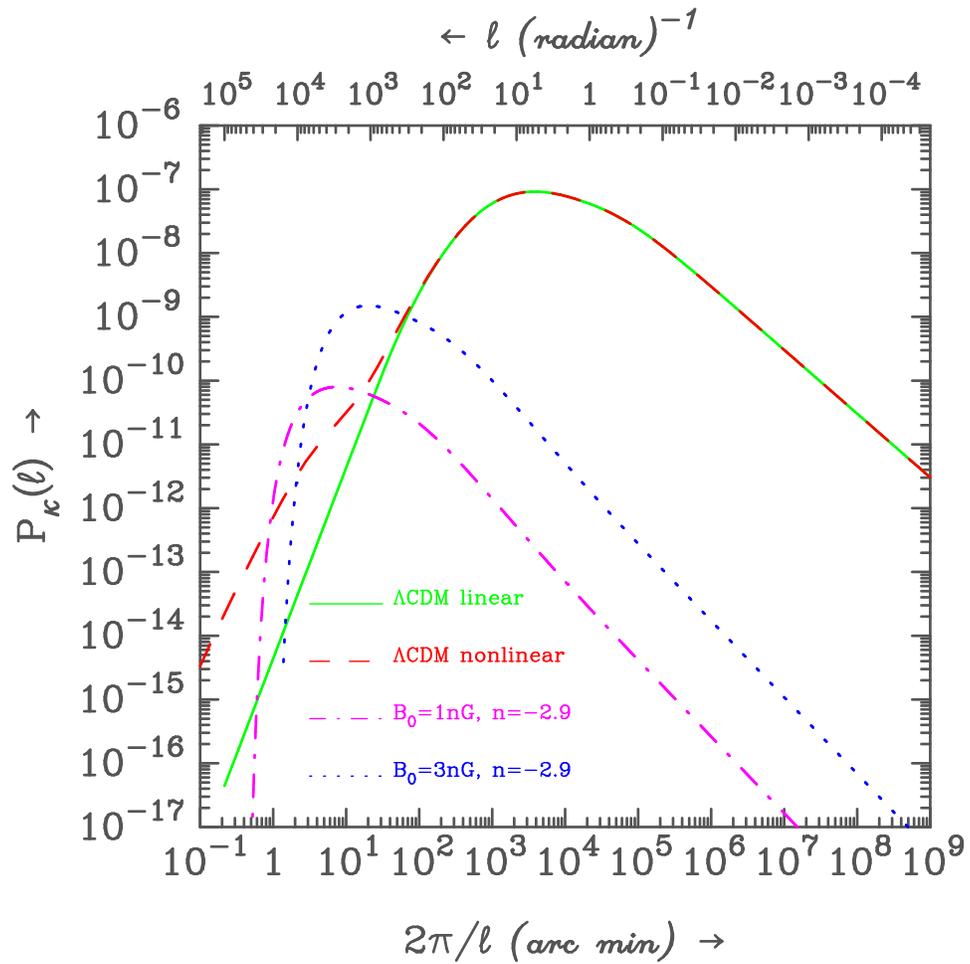}
}
   \caption[]{
Shear power spectra  for the magnetic and the $\Lambda$CDM models. 
}
\end{figure}
\clearpage
\vspace{0.1in}
\begin{figure}
   \centerline{
   \includegraphics[scale=0.7,angle=0]{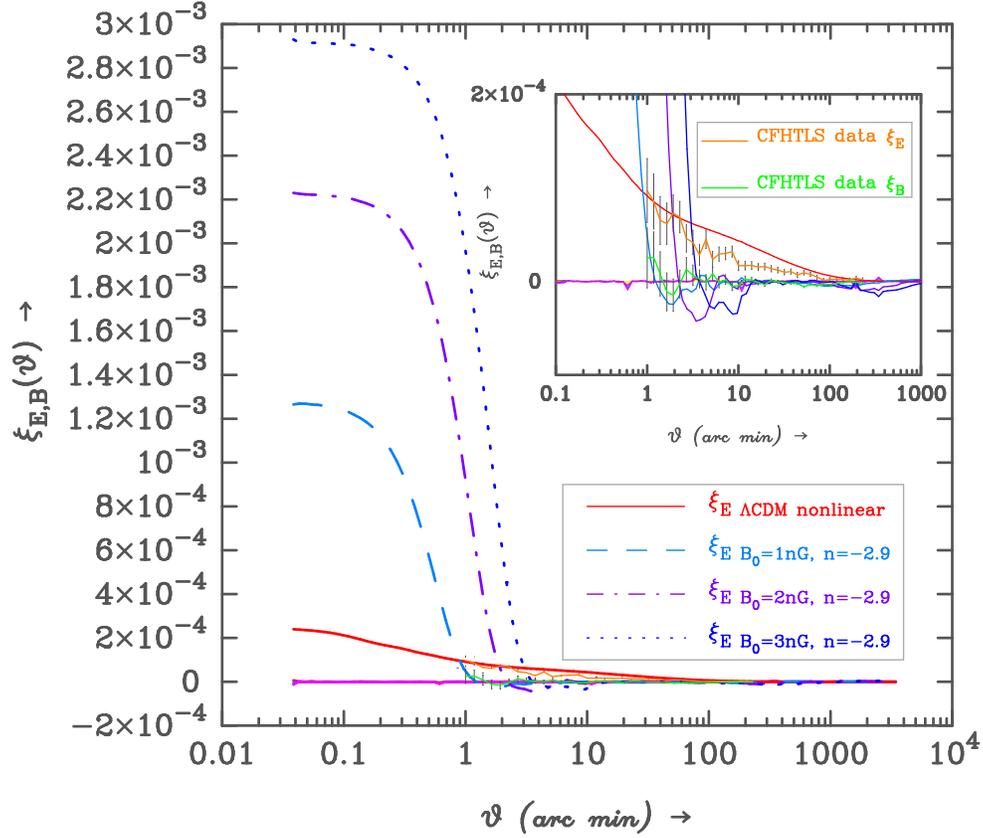}
}
   \caption[]{
Decomposed 2-point shear correlation functions $\xi_{E,B}$ for magnetic and non magnetic cases along with CFHT Legacy Survey data.  The inset magnifies  
the relevant curves and data points for a smaller range of  ordinate values. The solid (magenta) curves correspond to $\xi_B$. 
}
\end{figure}
\vspace{0.1in}

\vspace{0.1in}
\begin{figure}
   \centerline{
   \includegraphics[scale=0.7,angle=-90]{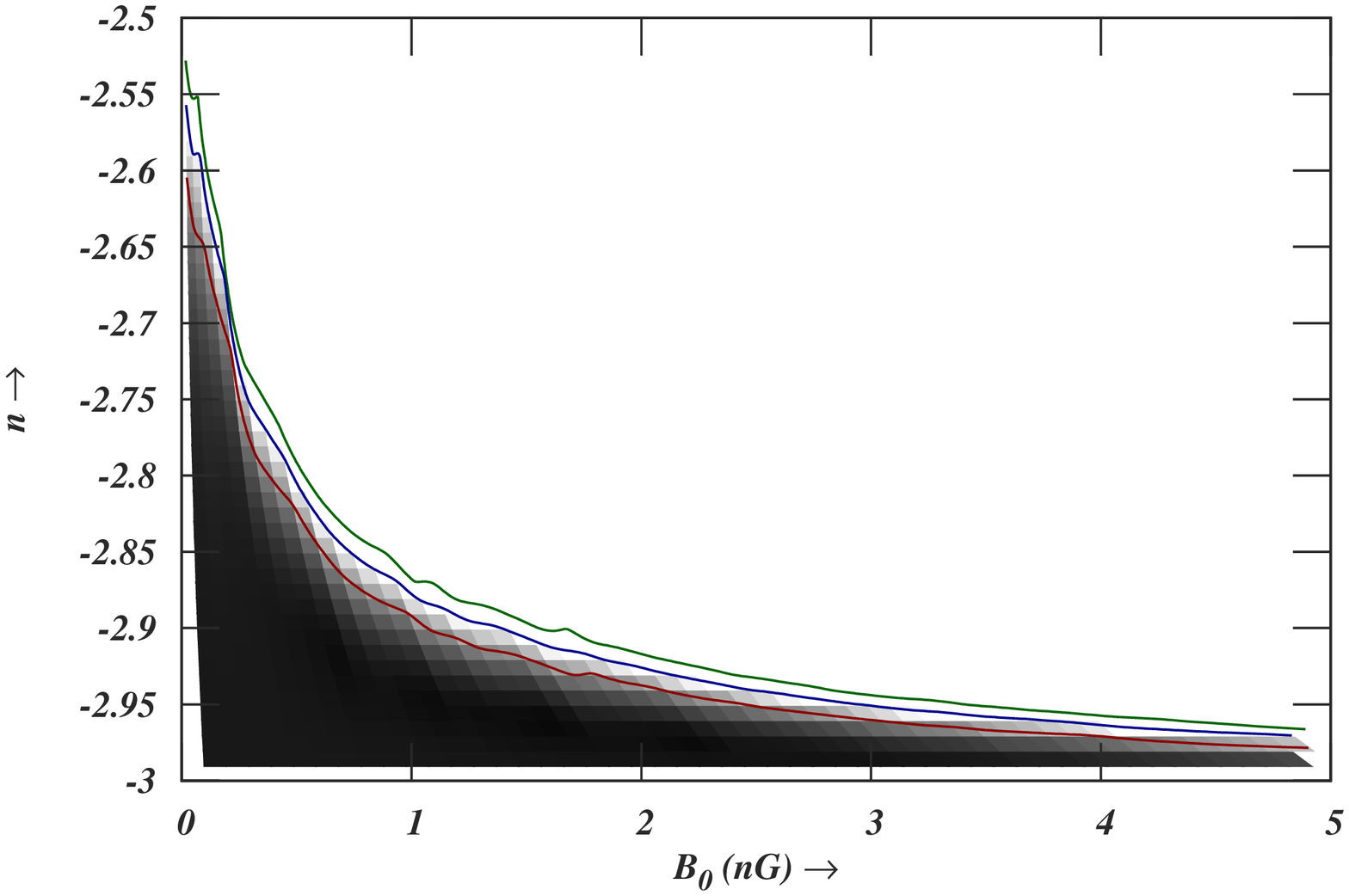}
}
\caption[]{
The figure shows the allowed region  in the ($B_0$,n) plane, 
 based on
the analysis of   ($({\xi_E})_B$ + $({\xi_E})_{\rm \Lambda CDM}$) against
 the CFHTLS data \citet{LFu08}. The shaded area is the 1-$\sigma$ allowed
region. 
The three  curves (from top to bottom) 
 are contours at 5$\sigma$, 3$\sigma$ and  1$\sigma$ level. 
}
\end{figure}
 
\section{Discussion}
Primordial magnetic fields leave their   signatures in a host
of observables in the universe. Their  impacts on CMBR temperature and 
polarization  anisotropies  have been extensively studied. \citet{Yamazaki10}
compute the allowed region in the $\{B_0,n\}$ plane by comparing the 
predictions of primordial magnetic field models with existing CMBR observations. 
Other constraints come from early formation of structures,  Faraday rotation
of CMBR polarization  \citep[e.g.][]{Tina2011} and reionization 
in the presence of magnetic fields \citet{SM2010}. 

In addition to the upper bounds on the magnetic field strength obtained 
by these observables, recent results suggests that there might be 
a lower bound of $\simeq 10^{-15} \, \rm G$ on the magnetic field strength 
(e.g.  \citet{Dolag10,NV2010,Tavecchio10,Taylor11}). This would suggest
that the magnetic field lies in the range  $10^{-15} < B_0 < \hbox{a few} \, 10^{-9} \, \rm G$. This range is still too large for a precise determination of the 
magnetic field strength. 

How do  our constraints (Figure~4) compare with the existing bounds on 
primordial magnetic fields? CMBR constraints (e.g. Figure~1 of \citet{Yamazaki10})
are stronger than our constraints for $n < -2.95$. For the entire range of 
spectral indices above this value, we obtain stronger upper limits on $B_0$. 
Our limits are comparable to bounds obtained from the formation of early structures, which also arise from excess power in the magnetic field-induced 
matter power spectrum (e.g. \citet{Tina2011}).  

Can primordial magnetic fields be detected in the Weak lensing data? As seen
in Figure~3, detection of excess power in the measurement of $\xi_E$ over what
is expected for the $\Lambda$CDM model, constrained well from other observations, could be interpreted as contribution 
from primordial magnetic fields. 

 The present data is noisy at the scales at 
which magnetic fields begin to make significant contribution, at least partly
owing to errors inherent in ground based measurements of shear, e.g. 
correction for point spread function, etc (e.g. Figure~4 
of \citep{schrabback2010}; a brief look at this figure might suggest
that their measurements would already put stronger constraints on magnetic 
field strength than presented here). Future, proposed space missions
such as  SNAP are likely to greatly improve the errors on these measurements. 
A comparison of Figure~4 of the white paper on weak lensing with
SNAP  \citep{albert2005} with the Figure~3 of this paper
suggests that SNAP would easily be able to probe sub-nano Gauss
 magnetic fields. 

 The magnetic field signal 
could be degenerate with the overall normalization  of the $\Lambda$CDM model
as measured by $\sigma_8$ ; WMAP 7-year data give $\sigma_8 = 0.801 \pm 0.030$ (\citep{Larson11}).  WMAP results are  in reasonable  agreement 
with the value of $\sigma_8$ as inferred by the weak lensing data. This
error is not sufficient to mimic the much larger  signal 
from  magnetic field strengths considered in this paper 
 (e.g. Figure~4 of\citep{schrabback2010}). However, a more careful analysis
will be needed to distinguish the error in $\sigma_8$
from the sub-nano Gauss magnetic fields.  
  
One uncertainty in our analysis is that the magnetic Jeans' scale, unlike
the thermal Jeans' scale which is well defined in linear perturbation
theory, is obtained within an approximation in which 
the backreaction of the magnetic field on the matter is not exactly  captured
\citep[e.g.][]{Kim96,Sethi05}.
 Even though our results capture 
qualitatively the impact of such a scale, there could be more power on 
sub-Jeans' scale which is lost owing to our approximation of the sharp k-cut
 off. As noted in section~2, the cut-off scale is the larger of the 
magnetic Jeans' length and the thermal Jeans' length. Magnetic field
dissipation can raise the temperature of the medium 
to $\simeq 10^4 \, \rm K$, thereby raising 
 thermal Jeans' length of the medium (Figure 4 
of \citet{SBK08} for a comparison between the two scales for different 
magnetic field strengths). For $B_0 \gtrsim  10^{-9} \, \rm G$, the magnetic  
Jeans' scale is the  larger of the two scales, as the maximum
temperature of the medium reached owing to this process doesn't exceed
$10^4 \, \rm K$.  In the more general case
also this would be true as photoionization of the medium by other sources, e.g.
the sources which could have cause reionization of the universe at
 $z\simeq 10$, 
also  results in comparable temperatures. For magnetic field strengths smaller
than considered in the paper, the cut-off scale is likely to be determined
by thermal Jeans' scale, caused  by the  photoionization of the medium by 
sources other than the magnetic field dissipation. 
Our approximation  allows us to
 identify important length and 
angular scales for our study (Figure~2 and~3). However, further work along 
these lines could extend our analysis by taking into account the physical 
effects of sub-magnetic Jeans' scales.

The analysis of Lyman-$\alpha$ forest in the redshift range $2 \lesssim z \lesssim 4$ is another powerful probe of the matter power spectrum of at small scales \citep[e.g.][]{croft02}.
Primordial magnetic field can  alter this interpretation in many ways: (a) more small scale power owing to magnetic field induced matter power spectrum (Figure~1), (b) dissipation of magnetic field can change the thermal state of Lyman-$\alpha$ clouds  \citep[e.g.][]{Sethi10,Sethi05}, (c) magnetic Jeans' length can reduce the  power at the smallest probable scale.  We hope to undertake this study in a future work.

\section*{Acknowledgments}
We thank an anonymous referee for useful comments which contributed to
improving the quality of the paper.

\end{document}